\documentclass[preprint,prb,nobibnotes,altaffilletter,amsmath,amssymb,amsfonts]{revtex4}

\usepackage{inputenc}
\inputencoding{latin1}
\usepackage{graphicx}
\usepackage{palatino} 
\usepackage{bm} 
\usepackage{color}
\usepackage[usenames,dvipsnames]{xcolor}

\usepackage{hyperref}

\newcommand{\eref}[1]{(\ref{#1})}

\definecolor{color2}{rgb}{1 0 0}
\definecolor{color3}{rgb}{0 0 1}
\definecolor{color4}{rgb}{0.5 0.6 1}
\definecolor{color5}{rgb}{0.8 0.3 0.9}
\definecolor{color6}{rgb}{0 1 1}
\definecolor{color7}{rgb}{1 0.0784 0.5765}
\definecolor{color9}{rgb}{0.4 1 0.1}
\definecolor{color10}{rgb}{1 0.6 0}

\begin{document}

\title{Generalized single-parameter aging tests and their application to glycerol}

\author{Lisa Anita Roed} \author{Tina Hecksher} \author{Jeppe C. Dyre} \author{Kristine Niss}
\affiliation{Glass and Time, IMFUFA, Department of
  Science and Environment, Roskilde University, Postbox 260, DK-4000 Roskilde,
  Denmark } \date{\today}

\begin{abstract}
  Physical aging of glycerol following temperature jumps is studied by dielectric
  spectroscopy at temperatures just below the glass
  transition temperature. The data are analyzed using two
  single-parameter aging tests developed by Hecksher \textit{et al.}
  [J. Chem. Phys. \textbf{142}, 241103 (2015)]. We generalize these
  tests to include jumps ending at different temperatures. Moreover,
  four times larger jumps than previously are studied. The
  single-parameter aging tests are here for the first time applied to a
  hydrogen-bonded liquid. We conclude that glycerol obeys
  single-parameter aging to a good approximation.
\end{abstract}

\maketitle{}
\section{Introduction}
Supercooled liquids are liquids cooled with a cooling rate fast enough
to avoid crystallization\cite{cotterill1985}. As
the glass transition is approached the liquid becomes more and more
viscous, and at the glass transition temperature ($T_g$) the molecules
effectively stop moving on the experimental time
scale\cite{brawer1985}. Supercooled liquids are often studied using
linear response experiments as for example dielectric spectroscopy,
where a change in the electric field (a perturbation) leads to a
change in polarization (a response). The perturbation is in this case usually small in order to ensure that the response depends linearly on the
applied perturbation, and the measured response function is connected
to the dynamics in equilibrium via the fluctuation dissipation
theorem.  In contrast to linear response experiments, aging is the
response following a large perturbation, where the liquid is brought
out of equilibrium, and the study of aging is a study of how the
properties of a system change over time as it relaxes toward
equilibrium\cite{scherer1986}. Aging is highly non-linear since the relaxation rate itself changes
as the liquid equilibrates.  Most aging studies, including the work
presented in this paper, are studies of how the system responds to a
change in temperature. The experiments are performed close to and
below the glass transition temperature, where aging takes place on
time scales that are slow enough to be measurable, yet fast enough to follow
all (or a substantial part) of the relaxation towards
equilibrium. Close to the glass transition temperature, the rate of
the structural relaxation depends strongly on temperature, and aging is non-linear even for fairly small
temperature steps. For a jump down in temperature the structural
relaxation becomes slower as time evolves, whereas for an jump up in
temperature the rate of structural relaxation increases as time
evolves.

The study of aging has a very practical root. The out-of-equilibrium
nature of glasses has the consequence that the properties of all
glassy materials age, even if it is sometimes very slow.
The standard formalism for describing this was set out by
Narayanaswamy, an engineer at Ford Motor Company who needed a method
for predicting how the frozen-in stresses in a wind-shield depend on
the glass' thermal history \cite{narayanaswamy1971}. Inspired by Tool, Narayanaswamy introduced a single-parameter aging
assumption in terms of a fictive temperature\cite{hecksher2015}. Today this is
referred to as the Tool-Narayanaswamy (TN) formalism. Besides the
obvious importance for application of glasses, aging experiments
have the potential to yield new information about fundamental
outstanding questions regarding the equilibrium relaxation (e.g. Ref. \onlinecite{niss2017}).

The analysis in this paper is based on the
TN aging formalism
\cite{tool1946,narayanaswamy1971,scherer1986,dyre2015}. The
TN-formalism interprets aging in terms of a material time $\xi$. The
material time may be thought of as a time measured on a clock with a
clock rate that changes as the system ages. The material
time is defined from the clock rate $\gamma(t)$
by\cite{narayanaswamy1971,dyre2015}
\begin{equation}
 d\xi=\gamma(t)dt.
\end{equation}
Narayanaswamy's idea was that linearity of the response is restored using the material time implying that the non-linear aging can be described using a linear convolution integral. The central hypothesis of single-parameter aging is that the clock-rate, $\gamma(t)$, depends only on one structural
parameter, and that this parameter also controls the measured
quantity.

Hecksher \textit{et al}.\ 2015\cite{hecksher2015} derived two tests of the single-parameter assumption, which are used and further developed below. The conceptual starting point of the tests is an ``ideal aging experiment'' in which
aging is measured from equilibrium to a new equilibrium after an
``instantaneous'' jump in temperature. In Sec. \ref{experimental}, we
discuss how to perform an experiment that comes close to the ideal
aging experiment. The tests from 2015 have the advantage that it is
not necessary to calculate the material time explicitly in order to
test for single-parameter aging. The tests are in these
respects similar to tests we developed in 2010
\cite{hecksher2010} and 2017 \cite{niss2017}. The single-parameter aging tests can only be applied for simple jumps in temperature, whereas other applications of the TN formalism may be used on more advanced thermal histories. However, usually other applications of the TN formalism assumes that the shape of the relaxation curve can be described by, e.g., a streched exponential and some functional form for how the clock rate depends on temperature and fictive temperature $\gamma=\gamma(T,T_f)$.

The unique feature of
one of the 2015-tests is that if a liquid obeys single-parameter aging,
then one relaxation curve can be used to predict other relaxation curves. In particular, this implies that it
is possible to predict linear relaxation curves from the non-linear
curves without evaluating the material time. The linear response to a temperature jump is not easy to
measure because it requires a very small temperature change. In this
paper we extend the tests and procedures from Hecksher \emph{et al.}
2015 to work also for jumps with different final temperature. 

Hecksher \textit{et al.}\cite{hecksher2015} concluded that the three
van der Waals liquids investigated all conform to single-parameter
aging to a good approximation. It is of interest to test whether
hydrogen-bonded liquids also exhibit single-parameter aging. The
difference between van der Waals bonded liquids and hydrogen-bonded
liquids is, in particular, important in relation to the isomorph
theory\cite{gnan2009,dyre2014}, which predicts that van der Waals
liquids have simple behavior along isochrones, e.g., isochronal superposition\cite{tolle2001,roland2003,pawlus2003,ngai2005,roland2008}, whereas no predictions
are given for hydrogen-bonded liquids. Investigations of both types of liquids are particularly
relevant following the recent development of isomorph theory of physical
aging\cite{dyre2018}. 

The TN-formalism was originally developed
for oxide glasses, which are covalently bonded systems, i.e.,
have strongly directional bonds. With that in mind it would not be
surprising to find that the TN-formalism works also for hydrogen-bonding systems. On
the other hand, the single-parameter tests we use are performed on high-precision data and the tests are developed to have
no free fitting parameters. So far we have only performed these high precision parameter-free tests on van der
Waals bonded liquids \cite{hecksher2010,hecksher2015,niss2017}, but in
this paper we present data and tests on the hydrogen bonding liquid,
glycerol. Physical aging of glycerol has been studied before, e.g., in Refs.\  \onlinecite{moynihan1991,fujimori1992,miller1997,miller1997b,simon1997,lunkenheimer2005}, with some of them showing that glycerol comply to the TN model. Other alcohols were also shown to comply to the TN model \cite{wang2008}. All these investigations, however, involve one or more free parameters.

Glycerol (propane-1,2,3-triol) is a small molecule with three hydroxyl
groups. It is the molecular liquid most often studied in
glassy dynamics, e.g.  Refs.\
\onlinecite{Gibson1923,Wang1971,Schiener1996,Berthier2005,Pezeril2009,Albert2016},
and it is of importance in technology, notably due to its
cryoprotectant properties \cite{Salt1958,Dashnau2006,Li2008}. For
these reasons glycerol is sometimes referred to as the archetypical glass
former. Yet, it is also known that glycerol supports the formation of a 3D
hydrogen-bonded network penetrating the bulk liquid\cite{Towey2011},
and it was recently shown that glycerol exihibits a low-frequency mode in the mechanical relaxation spectra
\cite{Jensen2018}. Thus there is dynamics on a slower time scale
than the main relaxation -- possibly due to the hydrogen bonding
network. This could influence the structural state of the liquid
and thereby the number of parameters involved in physical
aging.

Section \ref{experimental} gives an overview over the experimental
details. Section \ref{measurements} presents the data and initial
data treatment. Generalized single-parameter aging tests are derived
in Sec. \ref{singleparameter}, where also the results of the tests are
shown. Section \ref{discussion} discusses the generalized
single-parameter aging tests and their ability to predict jumps, and our
results are discussed in light of the isomorph theory.

\section{Experimental protocol and details}
\label{experimental}

\subsection{General considerations on the protocol}
We ideally want to perform a temperature jump instantaneously. Since
this is not possible, we instead aim for the time it takes to change
the temperature and reach thermal equilibrium throughout the sample,
to be much smaller than the structural relaxation time of the
liquid. In this case we may assume that no structural changes takes place during the temperature jump, and that the liquid
experiences the jump as ``instantaneous''. 

In order to study a full aging curve, it is important to apply the perturbation at
the right temperature. If the temperature is too far below $T_g$, the
liquid will not reach equilibrium on the experimental time scale, because of the strong temperature dependence of the structural relaxation time. On the other
hand, if the perturbation is applied at a temperature too far above $T_g$,
the liquid will come into equilibrium too fast for the aging to be measureble. This, of course, depends on how
fast one can measure. If the relaxation time is close to the time it
takes to make a measurement, structural relaxation will occur during
the measurement. This leaves only a small temperature interval where
conditions are optimal. 

With our equiliment we can perform the temperature jump in a few seconds. The starting temperature is a few
degrees below the conventional glass transition temperature (at which the
relaxation time is 100 s\cite{debenedetti1996}). At the temperatures
studied the relaxation time of the liquid is between 200 s and 23 hours. Table \ref{tablejumps} shows relaxation time for each temperature along with the annealing time for each jump which ranges from a couple of hours to more than 20 days. 
\begin{table}
	\begin{tabular}{|l|l|l|l|}
		\hline
		End temp. & Jump & Annealing time & $\tau_{eq}$ at end temp.\\
		\hline
		184 K & \textcolor{color2}{180 K $\rightarrow$ 184 K} & $\log(t/s)=3.7$ $\approx 1.4$ hours &$\log(\tau_{eq}/s)=2.3$ $\approx 3$ min  \\
		& \textcolor{color10}{176 K $\rightarrow$ 184 K} & $\log(t/s)=3.9$ $\approx 2.2$ hours& \\%$7.9 \cdot 10^3$ s
		\hline
		180 K & \textcolor{color3}{184 K $\rightarrow$ 180 K} & $\log(t/s)=4.8$ $\approx 18$ hours & $\log(\tau_{eq}/s)=3.6$ $\approx 66$ min\\
		& \textcolor{color7}{178 K $\rightarrow$ 180 K} & $\log(t/s)=4.8$ $\approx 18$ hours & \\
		& \textcolor{color5}{176 K $\rightarrow$ 180 K} & $\log(t/s)=5.0$ $\approx 30$ hours& \\
		\hline
		178 K & \textcolor{color6}{180 K $\rightarrow$ 178 K} & $\log(t/s)=5.5$ $\approx 90$ hours& $\log(\tau_{eq}/s)=4.3$ $\approx 5.5$ hours\\
		\hline
		176 K & \textcolor{color9}{184 K $\rightarrow$ 176 K} & $\log(t/s)=6.3$ $\approx 23$ days& $\log(\tau_{eq}/s)=4.9$ $\approx 23$ hours\\
		& \textcolor{color4}{180 K $\rightarrow$ 176 K} & $\log(t/s)=6.2$ $\approx 18$ days& \\
		\hline
	\end{tabular}
	\caption{Times. The annealing time for each jump ranges from a couple of hours to more than 20 days. The relaxation time at the end temperature $\tau_{eq}$ is found from Eq. \eref{predict_t2_all}.}
	\label{tablejumps}
\end{table}

\subsection{Temperature control}
In order to change the temperature fast, we use a specially designed
microregulator as sample
cell\cite{hecksher2010,niss2012,hecksher2015,niss2017}. The
microregulator is placed in a main cryostat\cite{igarashi2008b}. The microregulator keeps temperature fluctuations below 100
$\mu K$ and the characteristic thermal equilibrium time is 2~s. More
details are given in Ref. \onlinecite{hecksher2010}. In the studied
temperature range the time it takes to change temperature is below the
relaxation time of the liquid. We can therefore assume that little
structural relaxation takes place in the liquid during the temperature
change (with the possible exception of jumps starting at 184 K, where
the relaxation time is only 100 times longer that the thermal
equilibrium time).

\subsection{Permitivity -- the monitored property}
\begin{figure}
\begin{minipage}{0.32\textwidth}
\includegraphics[width=0.95\textwidth]{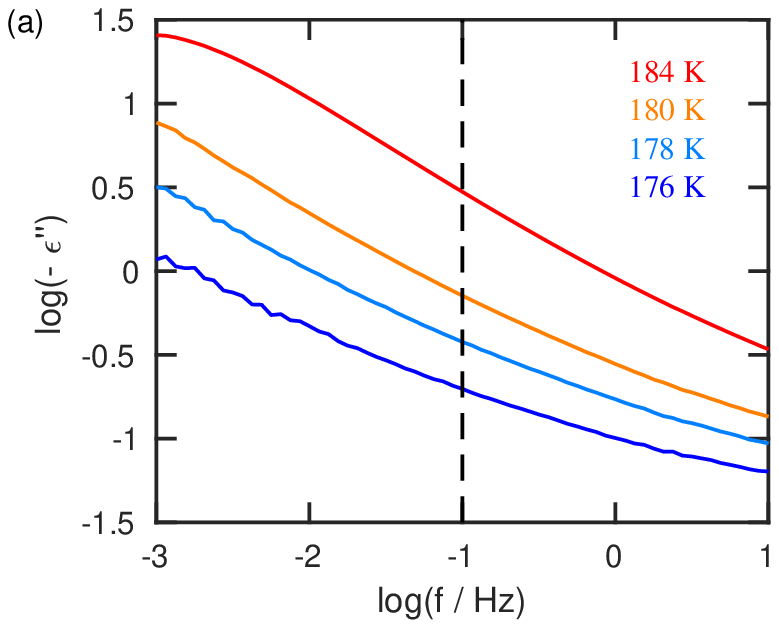}
\end{minipage}
  \begin{minipage}{0.32\textwidth}
  \includegraphics[width=0.95\textwidth]{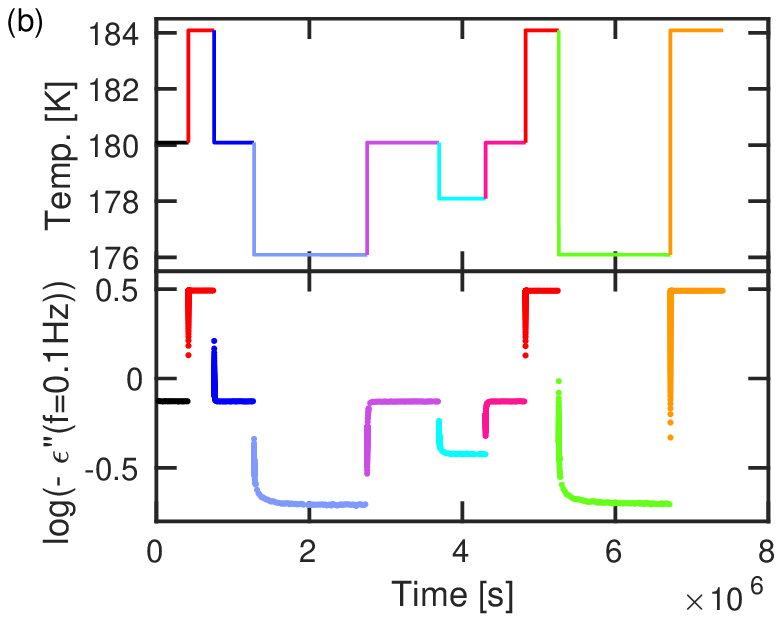}
\end{minipage}
\begin{minipage}{0.32\textwidth}
  \includegraphics[width=0.95\textwidth]{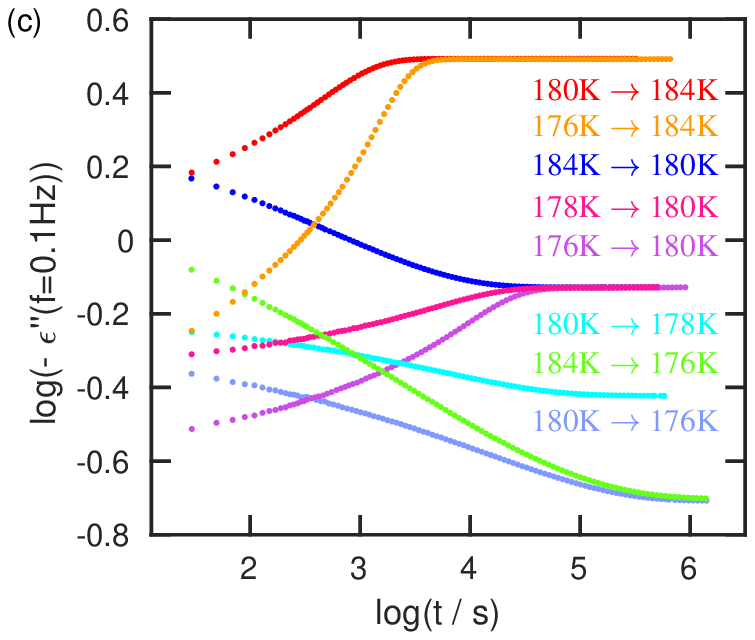}
\end{minipage}
  \caption{Raw data. (a) The dielectric loss as a function of frequency at equilibrium at the temperatures the jumps are performed between. The dashed black line marks the frequency 0.1 Hz used for monitoring aging. (b) The temperature and the measured quantity ($\log(-\varepsilon''(f=0.1 \text{Hz}))$) as functions of time. (c) The jumps on a logarithmic time scale. The data have been averaged in order to reduce noise as described in the text.}
  \label{figure1}
\end{figure}

An aging experiment monitors how a selected quantity evolves over
time. In this work we monitor the dielectric permitivity at a fixed
frequency. The dielectric signal is a useful probe because it can be
measured with high precision and has therefore been used to monitor
aging in several earlier works e.g. Schlosser and Sch\"onhals
\cite{schlosser1991}, Loidl and Lunkenheimer \textit{et
  al.}\cite{schneider2000,lunkenheimer2005,lunkenheimer2006,wehn2007},
Richert \textit{et al.}\cite{richert2013}, Alegr{\'i}a \textit{et
  al.}\cite{alegria1995,alegria1997,goitiandia2004}, and Cangialosi
\textit{et al.}\cite{cangialosi2005}. We use dielectric spectroscopy
with a parallel plate capacitor with diameter 10 mm and liquid layer
of 50 $\mu m$. The electrical measurement
equipment is described in
Ref. \onlinecite{igarashi2008a}. In order to perform the measurements fast,
a single measuring frequency is chosen, at which the dielectric loss is
measured as a function of time. The frequency is
chosen to be on the high-frequency flank of the $\alpha$-peak. Figure \ref{figure1} (a) shows the dielectric equilibrium
spectra at the temperatures where the jumps are performed, marking the
frequency used for monitoring 0.1 Hz, which corresponds to a acquisition
time of 20 s. For most measurements, we can assume that no structural
changes take place in the liquid during the measurement; however, for
the measurements starting at the highest temperature (184 K) where the
equilibrium relaxation time is just 200 s, some structural relaxation may occur during
the first few measurements.

\subsection{Sample}
Glycerol was acquired from Fluka and placed in an
desiccator for 18 h before the experiment. In order to prevent the
liquid taking up water from the surrounding air, the cryostat was
heated to 300 K before use. It was then flushed with liquid nitrogen,
after which the microregulator with sample was placed in the cryostat
and the temperature was lowered to 245 K (well below the melting
temperature of water).

\section{Measurements and initial data analysis}
\label{measurements}

\subsection{Protocol and raw data}
The measuring protocol is the following: The liquid is brought into equilibrium at the starting temperature 180 K (a few degrees below $T_g\approx 185$ K). We spent 35 days ($\log(t_{anneal} / s)=6.5$) for annealing the sample; equilibrium spectra were taken at different temperatures while cooling to estimate the right starting temperature. A jump in temperature is performed, and the liquid is monitored until equilibrium is established. A new jump is then performed. During the jumps the dielectric loss at the measuring frequency is monitored as a function of time ($\log(-\varepsilon''(f=0.1 \text{Hz},t))$). The jump magnitudes are 2 K, 4 K, and 8 K.

The temperature protocol and the raw data of the aging measurements are seen in Fig. \ref{figure1} (b), where each color represents a jump. Throughout the paper, reddish color tones are used for up jumps and bluish color tones are used for down jumps. The jump from 180 K to 184 K is performed twice, which served as a check of reproducibility, but only the first jump is used in the further data treatment. The total time of the measurement series is several months. The data are available on Glass and Time's data repository (http://glass.ruc.dk).

\subsection{Initial data analysis}
Figure \ref{figure1} (c) shows the data on a logarithmic time scale. The data are here averaged in order to reduce noise, which is done on a logarithmic scale by proceeding as follows. The first 15 raw data points are kept, then the $\log(time)$-scale are divided into 90 equal sized intervals over which the raw data are averaged. Figure \ref{figure1} (c) demonstrates that different jumps ending at the same temperature reach the same equilibrium value. It is on the other hand not clear in this figure that jumps starting at the same temperature have the same initial value. This is due to the instantanous contribution to the aging. Note that same size up and down jumps are not symmetric, which is due to the non-linearity seen even for small jumps. 

\begin{figure}
\begin{minipage}{0.45\textwidth}
	\includegraphics[width=0.95\textwidth]{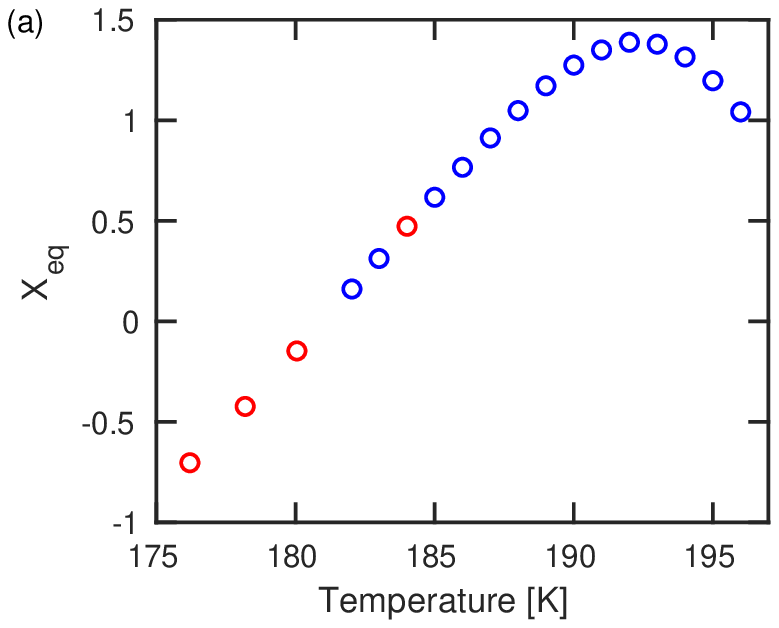}
\end{minipage}
\begin{minipage}{0.45\textwidth}
	\includegraphics[width=0.95\textwidth]{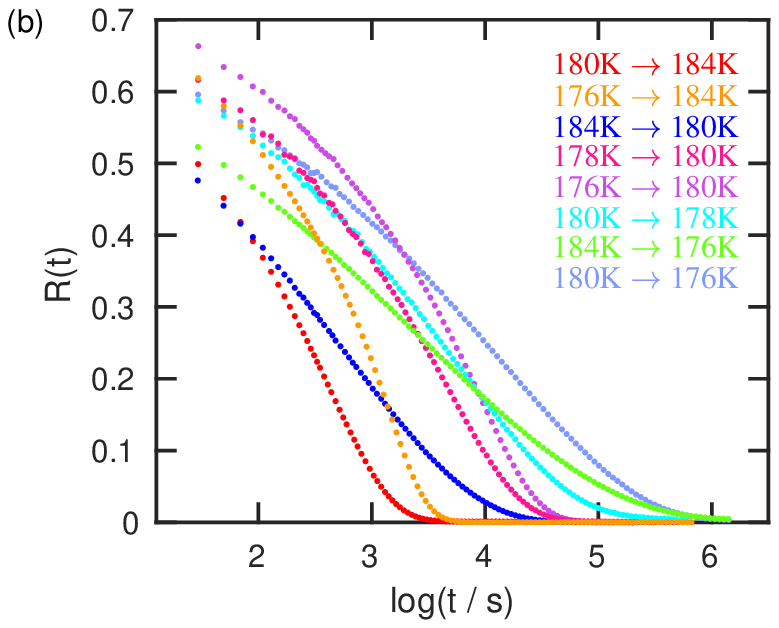}
\end{minipage}
  \caption{Quantities used in the data analysis to perform the
    single-parameter tests. (a) The measured quantity
    ($\log(-\varepsilon''(f=0.1 \text{Hz},t))$) at equilibrium at the
    final temperature ($X_{eq}$) as a function of temperature. The red
    circles indicate the four temperatures between which jumps are
    performed. The peak appearing between 190 K and 195 K is due to
    the $\alpha$-peak moving across the measuring frequency. (b) The
    normalized relaxation functions as functions of time for all
    jumps.}
  \label{figure2}
\end{figure}

For the further data treatment some notation is now introduced. The jump is performed from the starting temperature $T_{start}$ and ends at the temperature $T_{end}$. The measured time-dependent quantity is denoted $X(t)=\log(-\varepsilon''(f=0.1 \text{Hz},t))$. The jump starts at $t=0$ and the starting value of the measured quantity $X(0)$ is the equilibrium value of the measured quantity at $T_{start}$. When the liquid is in equilibrium at $T_{end}$ as $t\rightarrow \infty$, the measured quantity is $X_{eq}$. Figure \ref{figure2} (a) shows $X_{eq}$ as a function of temperature. The time-dependent distance to equilibrium at $T_{end}$ is denoted $\Delta X(t)=X(t)-X_{eq}$. $\Delta X(0)$ is the total change of the measured quantity from $T_{start}$ to $T_{end}$. The definition of $\Delta X(t)$ is a generalization of other concepts that quantify the departure from equilibrium, e.g., Kovacs' expression in terms of volume $\delta(t)=(V(t)-V_{\infty})/V_{\infty}$\cite{kovacs1963}.

The normalized relaxation function $R(t)$ is defined as the
time-dependent distance to equilibrium over the overall change from
start to end of the measured
quantity\cite{narayanaswamy1971,scherer1986}
\begin{equation}
\label{def_R}
 R(t)=\frac{\Delta X(t)}{\Delta X(0)}
\end{equation}
It is seen that $R(0)=1$ and that $R(t)\rightarrow 0$ as
$t\rightarrow\infty$. Figure \ref{figure2} (b) shows $R(t)$ for all
jumps. According to the TN formalism $R(t)$ is a unique function of
the material time $\xi$\cite{scherer1986}.

\section{Generalized single-parameter aging tests}
\label{singleparameter}
Two single-parameter aging tests were derived by Hecksher \textit{et
  al.} 2015\cite{hecksher2015} for jumps ending at the same
temperature. In this investigation we study jumps to different
temperatures, and the single-parameter aging tests are therefore
generalized below.

According to the single-parameter assumption, both the measured
quantity $X(t)$ and the clock rate $\gamma(t)$ are controlled by the
same parameter $Q(t)$. Hecksher \textit{et al.}\cite{hecksher2015}
assumed that the temperature jumps are so small that it is reasonable to
Taylor expand both $X(t)$ and $\ln\gamma(t)$ to the first order in
$Q(t)$: $\Delta X(t)\cong c_1\Delta Q(t)$ and
$\Delta \ln\gamma(t)\cong c_2\Delta Q(t)$, where $c_1$ and $c_2$ are
constants and $\Delta X(t)$, $\Delta Q(t)$, and $\Delta \ln\gamma(t)$ defines the time-dependent distance to equilibrium for respectively $X(t)$, $Q(t)$, and $\ln\gamma(t)$. Eliminating $\Delta Q(t)$ in the equations gives
$\ln\gamma(t)=\ln\gamma_{eq}+\Delta X(t)/X_{const}$, where
$X_{const}=c_1/c_2$. Using the definition of $R(t)$ (Eq. \eref{def_R})
gives
\begin{equation}
  \label{gamma}
  \gamma(t)=\gamma_{eq}\exp \left(\frac{\Delta
      X(0)}{X_{const}}R(t)\right) 
\end{equation}
As expected $\gamma(t)\rightarrow \gamma_{eq}$ as $t\rightarrow\infty$
because $R(t)\rightarrow 0$.  Hecksher \textit{et
  al.}\cite{hecksher2015} derived the following expression for the
time-derivative of $R$ from the TN-formalism
\begin{equation}
  \label{dRdt_F_gamma}
  \dot{R}=-F(R)\gamma(t)
\end{equation}
where $F(R)$ is a unique function of $R$, i.e., independent of start and
end temperature. Since $R(t)$ goes from 1 to 0, $\dot{R}<0$, and $F(R)>0$ per definition.. Inserting
the expression for $\gamma$ (Eq. \eref{gamma}) and rearranging leads
to
\begin{equation}
  \label{dotR}
  -\frac{\dot{R}}{\gamma_{eq}}\exp\left(-\frac{\Delta
    X(0)}{X_{const}}R(t)\right) = F(R)
\end{equation}
from which the two generalized tests are derived in the following
sections.

\subsection{Test 1 - Predicting general jumps from knowledge of a
  single jump}
\label{singleparametertest1}
If single-parameter aging is obeyed, it is possible to predict the
relaxation function for one jump from the relaxation function of
another jump. In the following we use subscripts 1 and 2 to
distinguish between the known (measured) relaxation function -- jump 1
-- and the predicted relaxation function -- jump 2. From one jump
relaxation function $R_1(t)$ and its inverse function $t_1(R)$, we
derive a method for determining $t_2(R)$ for a different jump $R_2(t)$
using Eq.\ (\ref{dotR}).

At times $t_{1}^{*}(R)$ and $t_{2}^{*}(R)$ where the value of the
relaxation functions are the same ($R=R_1=R_2$), Eq. (\ref{dotR})
implies
\begin{equation}\label{dotReq}
  -\frac{dR_1}{dt_{1}^{*}} \cdot
  \frac{1}{\gamma_{eq,1}}\cdot\exp\left(-\frac{\Delta 
      X_1(0)}{X_{const}}R_1(t_1^*)\right) =
  -\frac{dR_2}{dt_{2}^{*}} \cdot \frac{1}{\gamma_{eq,2}} \cdot
  \exp\left(-\frac{\Delta X_2(0)}{X_{const}}R_2(t_2^*)\right) \,.
\end{equation}
For time increments $dt_{1}^{*}$ and $dt_{2}^{*}$ leading to identical
changes $dR_1=dR_2$ we can write Eq.\ (\ref{dotReq}) using the premise
that $R_1(t_1^*) =R_2(t_2^*)$ to give
\begin{equation}
  dt_{2}^{*} = \frac{\gamma_{eq,1}} {\gamma_{eq,2}} \cdot \exp \left( 
    \frac{\Delta X_1(0) - \Delta X_2(0)}{X_{const}} R_1(t_1^{*}) 
  \right)dt_1^{*}\,.
\end{equation}
Integrating this leads to (assuming that both jumps are initiated at
time zero)
\begin{equation}
  \label{predict_t2_all}
  t_2 = \int_0^{t_2} dt_{2}^{*}
  =\frac{\gamma_{eq,1}}{\gamma_{eq,2}}\int_0^{t_1} 
  \exp \left( \frac{\Delta X_1(0)-\Delta X_2(0)}{X_{const}}
    R_1(t_1^*)\right)  dt_1^* \,,
\end{equation}
which determines $t_{2}(t_1)=t_2(t_1(R))=t_2(R)$, predicting the
inverse function $R_2(t)$.

In practice, we do not need to do an inversion. The procedure is
to transform a discrete set of measured data points, a time vector
${\bf t}_1=(t_1^1,t_1^2,\ldots,t_1^n)$ and the corresponding relaxation
vector ${\bf R}_1=(R_1^1,R_1^2,\ldots,R_1^n)$) to a new time vector
${\bf t}_2=(t_2^1,t_2^2,\ldots,t_2^n)$ corresponding to the measured
${\bf R}_1$ points. Plotting $({\bf t}_2,{\bf R}_1)$ should then coincide with $R_2$, which can be tested by a separate measurement.

For jumps ending at the same temperature Eq. \eref{predict_t2_all}
reduces to
\begin{equation}
  \label{predict_t2}
  t_{2} = \int_0^{t_1} \exp\left(\frac{\Delta X_1(0)-\Delta
      X_2(0)}{X_{const}} R_1(t_1^*)\right)  dt_1^*\,, 
\end{equation}
which is the equation used in Hecksher \textit{et al.}\cite{hecksher2015}.

The new generalized test has the disadvantage that one needs to know
the equilibrium clockrate $\gamma_{eq}$. By assuming an internal
clock\cite{hecksher2010,jakobsen2012}, i.e., that the clock rates from dielectric spectroscopy and structural relaxation are proportional, one may define the clock rate
at $T_{end}$ ($\gamma_{eq}$) as the dielectric inverse relaxation time
($\gamma_{eq}=1/\tau$), where the relaxation time is determined from
the maximum frequency at the $\alpha$-peak ($f_m$) as
$\tau=1/(2\pi f_{m})$. At the temperatures where jumps are performed,
we cannot determine $\tau$ directly, since the $\alpha$-peak is outside the
available frequency range (see Fig. \ref{figure1} (a)). One may find
$\gamma_{eq}$ using an extrapolation or determine $\gamma_{eq}$ in another
way. Below we let the constant $\frac{\gamma_{eq,1}}{\gamma_{eq,2}}$
in Eq. \eref{predict_t2_all} be a free parameter to find the
$\gamma_{eq}$ that gives the best prediction (which is very close to
the prediction using a Vogel-Fulcher-Tammann (VFT) extrapolation as shown in Fig. \ref{figure4}).

The constant $X_{const}$ was identified using two jumps to the same
temperature; the down jump 184 K to 180 K and the up jump 176 K to 180
K. A fit is made using Eq. \eref{predict_t2} to find the $X_{const}$
that gives the best prediction of one jump from the other. This gives
$X_{const}=0.16$, which was used for all jumps.

\begin{figure}
  \includegraphics[width=0.45\textwidth]{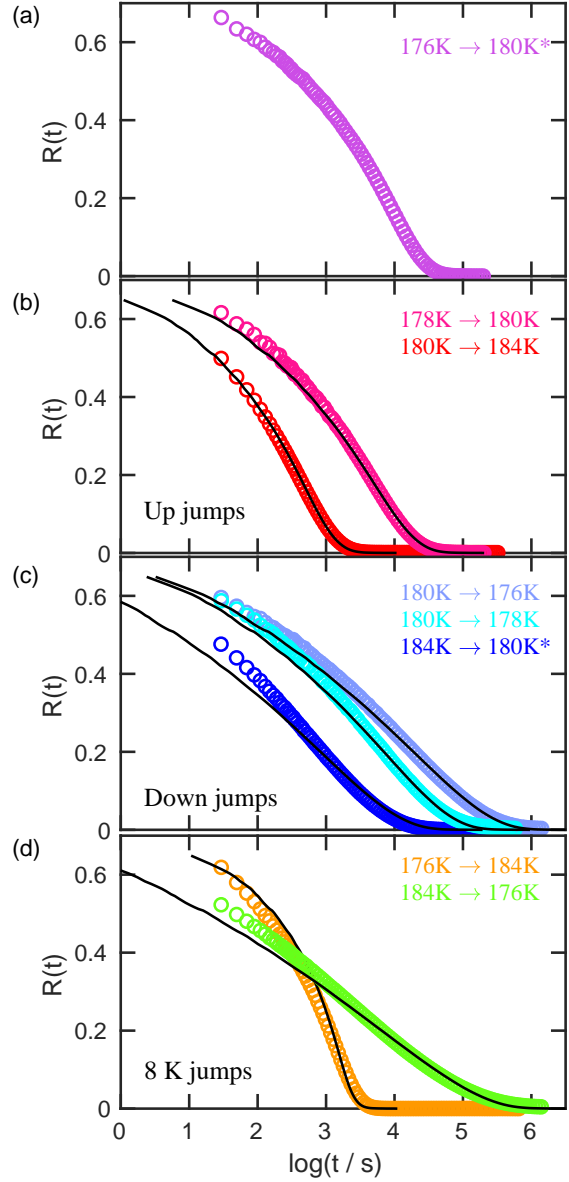}
  \caption{Test 1. Prediction of jumps based on the jump 176 K to 180 K (shown in (a)). In (b), (c), and (d) circles are data and black lines are predictions. The up jump 176 K to 180 K and the down jump 184 K to 180 K were used to find $X_{const}$ (indicated by * in legend), while $\gamma_{eq}(T)$ was a fitting parameter the consistency of which with equilibrium dynamics is checked in Fig. \ref{figure4}. The predictions are good, however, with deviations in particular at short times. It is also seen that the predictions are slightly worse for the largest jumps as well as for the down jumps starting at the highest temperatures. } 
  \label{figure3}
\end{figure}

\begin{figure}
	\includegraphics[width=0.45\textwidth]{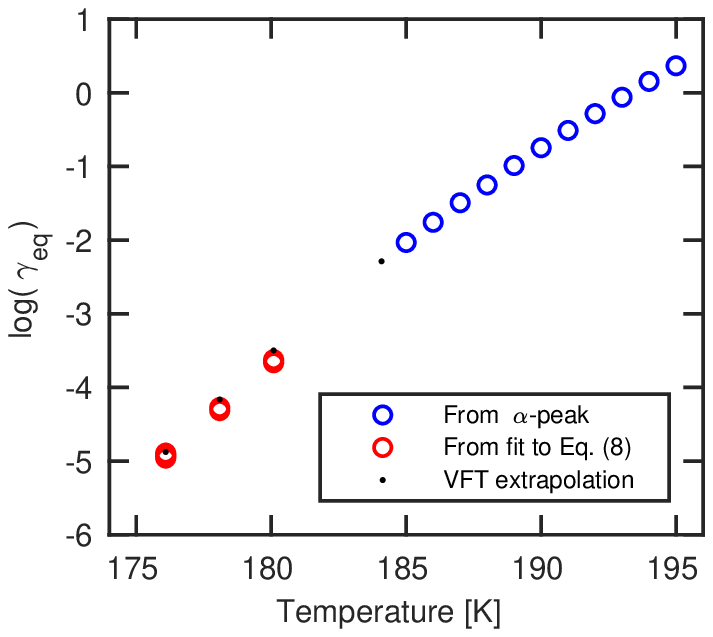}
  \caption{The clock rate at equilibrium at the end temperature
    ($\gamma_{eq}$) found from equilibrium spectra (blue circles). The
    red circles are determined from the free parameter
    $\gamma_{eq,1}/\gamma_{eq,2}$ in Eq. \eref{predict_t2_all}. Since
    we have two jumps to 180 K and to 176 K, this results in two values of
    $\gamma_{eq}$ at these temperatures. However, the difference is
    barely visible. The black dots are the predictions from the
    VFT-fit. }
	\label{figure4}
\end{figure}

The result of the test is shown in Fig. \ref{figure3}. The up jump 176
K to 180 K (Fig. \ref{figure3} (a)) is used to predict the other
jumps. 
The predictions are good, however with deviations
in particular at short times, similar to those observed for the van der Waals
liquids tested in Hecksher \textit{et al.
2015\cite{hecksher2015}.} 
The 8 K jumps have larger deviations than the
smaller jumps, however, still not more dramatic than results on van der
Waals liquids\cite{hecksher2015}. Since we use a first-order Taylor
expansion in the derivation of the test, it is not surprising that the predictions are slightly worse for large temperature jumps.

Note that the two down jumps starting at 184 K both have significant deviations at short times. This may be related to the relatively fast relaxation time at this starting temperature, and structural relaxation may have occured during the jump and during the first measurements making these jumps less ideal in the sense defined earlier.

Figure \ref{figure4} shows $\gamma_{eq}$ as a function of temperature found from the free parameter $\gamma_{eq,1}/\gamma_{eq,2}$ in Eq. \eref{predict_t2_all}. To determine $\gamma_{eq}$ for each temperature, we used a value of $\gamma_{eq}$ at 184 K found from a VFT-fit to higher temperature data. 
Different jumps to the different temperatures give a slightly different value, but this is barely visible in the figure. The values found are reasonable, but vary slighty from the predictions of $\gamma_{eq}$ from VFT. This supports the internal clock hypothesis, i.e., that the clock rates from dielectric spectroscopy and structural relaxation are proportional. This is in line with the result in Ref. \onlinecite{lunkenheimer2005}. The values of $\gamma_{eq}$ are given in Table \ref{tablegamma}.

\begin{table}
	\begin{tabular}{|r|r|r|r|}
          \hline
          Temperature & $\log(\gamma_{eq})$ from Eq. \eref{predict_t2_all} & Average $\log(\gamma_{eq})$ from Eq. \eref{predict_t2_all} & $\log(\gamma_{eq})$ from VFT\\
          \hline
          184 K &  & & -2.28 \\
          \hline
          180 K & -3.62 & -3.64 &-3.50 \\
                      & -3.67 & & \\
          \hline
          178 K & -4.27 & -4.29 & -4.16 \\
                      & -4.31  & & \\
          \hline
          176 K & -4.92 & -4.92 & -4.88 \\
                      & -4.96 & & \\
                      & -4.89 & & \\
                      & -4.93 & & \\
		\hline
	\end{tabular}
	\caption{The equilibrium clock rate at the end temperature
          $\gamma_{eq}$ determined from Eq. \eref{predict_t2_all}. To invert $\gamma_{eq,1}/\gamma_{eq,2}$ to $\gamma_{eq}$, we first used the value of $\gamma_{eq}$ at 184 K found from a VFT-fit to estimate $\gamma_{eq}$ at 180 K (resulting in two values because there are two jumps to 184 K). The estimated $\gamma_{eq}$ at 180 K are then used to determine $\gamma_{eq}$ at 178 K and 176 K. Different jumps to the same temperature give a slightly different value of $\gamma_{eq}$. Furthermore, values found from extrapolations of VFT-fit are given.}
	\label{tablegamma}
\end{table}

\subsection{Test 2 - Unique function of $R$}
Taking the logarithm of Eq. \eref{dotR} leads to (remember that
$\dot{R}<0$)
\begin{equation}
\label{LHS}
 \ln\left(-\frac{\dot{R}}{\gamma_{eq}}\right)-\frac{\Delta X(0)}{X_{const}}R=\ln(F(R))
\end{equation}
Since $F(R)$ is a unique function of $R$ if single-parameter aging applies, the left hand side of
Eq. \eref{LHS} (denoted LHS following Hecksher \textit{et al.}
2015\cite{hecksher2015}) is also a unique function of $R$. This means
that LHS plotted against $R$ for the different jumps collapse onto a
master curve if a single parameter controls both $X(t)$ and
$\gamma(t)$. $\gamma_{eq}$ found from Test 1 is used. For each
temperature the found values of $\gamma_{eq}$ is now averaged, so
that each temperature has a fixed $\gamma_{eq}$.

\begin{figure}
  \includegraphics[width=0.45\textwidth]{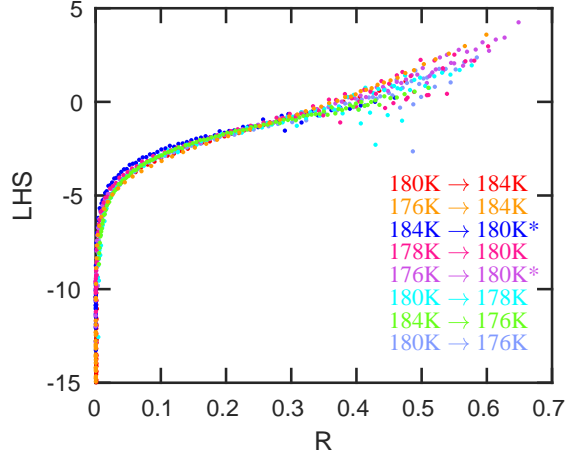}
  \caption{Test 2. The left hand side of Eq. \eref{LHS} (LHS) plotted against $R$ for all jumps. If the liquid obeys single-parameter aging, the data are predicted to collapse onto a master curve which is seen to apply to a good approximation.}
  \label{figure5}
\end{figure}

Figure \ref{figure5} shows that the data collapse onto a master curve. As also seen with test 1, deviations are observed in particular at short times (large $R$).

\section{Discussion \& conclusion}
\label{discussion}
\subsection{How to test the single-parameter assumption}
The data analysis emphasizes that measurements must be precise to allow for direct tests of single-parameter aging. It is important that the liquid is in equilibrium at the starting temperature. The liquid does not have to reach full equilibrium if one knows the correct $X_{eq}$; i.e., where it should end. Our analysis revealed that it is crucial to have the correct $X_{eq}$. Performing an up jump from the glassy state makes it possible to monitor a full relaxation curve where both the plateau in the beginning and the end are present. For this reason it is recommended to use such a jump for predicting other jumps.

Test 1, which is used to predict the curves, is the most sensitive test. By using test 1 it is possible to find $\gamma_{eq}$ at temperatures where the peak is not present in the dielectric equilibrium spectrum. If one knows $\gamma_{eq}$ (for example by an extrapolation from higher temperatures), an advantage of test 1 is that if a liquid obeys single-parameter aging, one can use Eq. \eref{predict_t2_all} to predict relaxation curves. It is necessary to perform two jumps (preferably to the same temperature) to identify $X_{const}$. Other jumps can then be predicted knowing the clock rate at the end temperature ($\gamma_{eq}$), the measured quantity at the starting temperature ($X(0)$), and the measured quantity at the end temperature ($X_{eq}$). $\Delta X(0)=X(0)-X_{eq}$ determines the shape of the relaxation, while $\gamma_{eq}$ determines the position of the relaxation. This is of course also limited to small jumps, and possible only in the same temperature range as the jumps are performed in.

\subsection{The scope of single-parameter aging}
Hecksher \textit{et al.} 2015\cite{hecksher2015} demonstrated
single-parameter aging for three different van der Waals liquids. In
the present investigation the tests have for the first time been applied to a
hydrogen-bonded liquid. The comparison of hydrogen-bonded liquids and
van der Waals bonded liquids is interesting in view of the isomorph
theory, which states that R-simple liquids have simple behavior along
isochrones\cite{gnan2009,dyre2014}. Van der Waals bonded liquids are
expected to be R-simple, whereas hydrogen-bonded liquids are
not. The isomorph theory was suggested a decade ago, and since then it has been
tested in a number of investigations,
e.g., Refs.\  \onlinecite{gundermann2011,roed2013,xiao2015,roed2015,adrjanowicz2016,hansen2018}. Earlier
investigations of density
scaling\cite{dreyfus2003,alba2004,casalini2004,roland2005} and
isochronal
superposition\cite{tolle2001,roland2003,pawlus2003,ngai2005,roland2008}
also follow the predictions of the isomorph theory. The understanding
of aging in computer simulation was connected to isomorph theory
already in 2010 \cite{Gnan2010}, and recently experimental results
\cite{niss2017} and aging theory \cite{dyre2018} were also analyzed
and developed in the framework of isomorph theory. Based on these
earlier works we expected that single-parameter aging might
not work for a hydrogen-bonding network-forming system like
glycerol. However, based on the tests presented above we find
that single-parameter aging works to a good approximation also for
glycerol. The quantitative agreement with the predictions from the
single-parameter aging found for glycerol is, in fact, similar to
that found for three van der Waals bonding liquids in
Ref. \onlinecite{hecksher2015}. This finding suggests that single-parameter
aging works for a wider range of systems than those complying to
isomorph theory, which also means that single-parameter aging needs to 
be understood in a more general theoretical framework. 

The tests work well also for the large jumps (4 K and 8 K), while they
previously were applied only to jumps of maximum 2
K\cite{hecksher2015}. Note that a jump of a certain temperature span corresponds to different spans in $\gamma_{eq}$ for different liquids. In particular the difference in fragility between hydrogen-bonded liquids and van der Waals bonded liquids means that temperature jumps of equal sizes corresponds to a larger change of $\gamma_{eq}$ for the van der Waals bonded liquids.  The 8 K jump of glycerol corresponds to $\Delta\log\gamma_{eq}\approx 2.6$, while $\Delta\log\gamma_{eq}\approx 0.3$ to 0.8 for the previously tested jumps on van der Waals liquids (the total span of $\gamma_{eq}$ for the used up and down jump is $\Delta\log\gamma_{eq}\approx 0.7$ to 1.5). This implies that also with respect to $\Delta\log\gamma_{eq}$ the jumps studied here are larger than tested before. For the 8 K jumps single-parameter aging may
begin to break down, but this is not surprising since a first-order
Taylor expansion is used in the derivation of the tests. For future
works higher-order Taylor expansions could be introduced into the tests.

\section{Acknowledgement}
L.\ A.\ R.\ and K.\ N.\ want to thank Independent Research Fund Denmark (Sapere Aude: Starting Grant) and Danish National Research Foundation's Grant No. DNRF61 for supporting this work. T.\ H.\ and J.\ C.\ D.\ are supported by the VILLUM Foundation's Matter grant
(16515).

%\bibliography{Aging_bib2}

\pagebreak

\end{document}